\documentclass[%
 reprint,
 amsmath,amssymb,a
 aps,
]{revtex4-2}

\usepackage{graphicx}
\usepackage{dcolumn}
\usepackage{bm}
\usepackage[colorlinks,linkcolor=blue,citecolor=blue]{hyperref}
\usepackage{lineno}
\usepackage{graphicx}

\usepackage{url}
\usepackage{ulem}
\usepackage{amsmath}
\usepackage{amsfonts}
\usepackage{textcomp}
\usepackage{subfigure}
\usepackage{verbatim}
\usepackage{rotating}

\def \ee   {e^+e^-}
\def \uu   {\mu^+\mu^-}

\def \jpsi {J/\psi}

\begin{document}

\title{\boldmath Search for a muonphilic scalar $X_{0}$ or vector $X_{1}$ via $J/\psi\to\mu^+\mu^-+\rm{invisible}$ decays at BESIII}

\author{
M.~Ablikim$^{1}$, M.~N.~Achasov$^{4,b}$, P.~Adlarson$^{75}$, O.~Afedulidis$^{3}$, X.~C.~Ai$^{80}$, R.~Aliberti$^{35}$, A.~Amoroso$^{74A,74C}$, M.~R.~An$^{39}$, Q.~An$^{71,58}$, Y.~Bai$^{57}$
% ...
\\
\vspace{0.2cm}
(BESIII Collaboration)\\
\vspace{0.2cm} {\it
$^{1}$ Institute of High Energy Physics, Beijing 100049, People's Republic of China\\
% ...
$^{3}$ Bochum Ruhr-University, D-44780 Bochum, Germany\\
$^{4}$ Budker Institute of Nuclear Physics SB RAS (BINP), Novosibirsk 630090, Russia\\
% ...
$^{35}$ Johannes Gutenberg University of Mainz, Johann-Joachim-Becher-Weg 45, D-55099 Mainz, Germany\\
% ...
$^{39}$ Liaoning Normal University, Dalian 116029, People's Republic of China\\
% ...
$^{57}$ Southeast University, Nanjing 211100, People's Republic of China\\
$^{58}$ State Key Laboratory of Particle Detection and Electronics, Beijing 100049, Hefei 230026, People's Republic of China\\
% ...
$^{71}$ University of Science and Technology of China, Hefei 230026, People's Republic of China\\
% ...
$^{74}$ University of Turin and INFN, (A)University of Turin, I-10125, Turin, Italy; (B)University of Eastern Piedmont, I-15121, Alessandria, Italy; (C)INFN, I-10125, Turin, Italy\\
$^{75}$ Uppsala University, Box 516, SE-75120 Uppsala, Sweden\\
% ...
$^{80}$ Zhengzhou University, Zhengzhou 450001, People's Republic of China\\
\vspace{0.2cm}
% ...
$^{b}$ Also at the Novosibirsk State University, Novosibirsk, 630090, Russia\\
% ...
}
\vspace{0.4cm}
}


\begin{abstract}
A light scalar $X_{0}$ or vector $X_{1}$ particles have been introduced as a possible explanation for the $(g-2)_{\mu}$ anomaly and dark matter phenomena. 
 Using $(8.998\pm 0.039)\times10^9$ $\jpsi $ events collected by the BESIII detector, we search for a light muon philic scalar $X_{0}$ or vector $X_{1}$ in the processes $\jpsi\to\uu X_{0,1}$ with $X_{0,1}$ invisible decays. 
No obvious signal is found, and the upper limits on the coupling $g_{0,1}'$ between the muon and the $X_{0,1}$ particles are set to be between $1.1\times10^{-3}$ and $1.0\times10^{-2}$ for the $X_{0,1}$ mass in the range of  $1<M(X_{0,1})<1000$~MeV$/c^2$ at 90$\%$ credibility level.
\end{abstract}

  

\oddsidemargin  -0.2cm
\evensidemargin -0.2cm
\maketitle

The Standard Model (SM) of particle physics has achieved remarkable successes as a highly predictive theory of fundamental particles and their interactions. Nonetheless, the SM is generally considered incomplete since it is unable to explain several important questions, anomalies, and phenomena~\cite{Crivellin:2021sff,Fox:2022tzz,Chen:2021fcb}. 
One of the possible experimental evidences of physics beyond the SM is the persistent  discrepancy of more than $3\sigma$ between the experimental observation and the SM prediction of the muon anomalous magnetic moment $(g-2)_{\mu}$~\cite{Muong-2:2006rrc,Muong-2:2021ojo,Muong-2:2023cdq}. 

Extra $U(1)$ groups have been added as minimal extensions to the SM to study new physics~\cite{Pospelov:2008zw,Bauer:2018onh}. One of the notable extensions of the SM gauge group is the anomaly-free gauged $U(1)_{L{\mu}-L{\tau}}$ model~\cite{He:1990pn,Foot:1990mn,Foot:1994vd}. This model introduces a new massive vector boson $X_{1}$, which only couples to the second and third generations of leptons ($\mu,\nu_{\mu},\tau,\nu_{\tau}$) with the coupling strength $g_{1}'$.
The $X_1$ can contribute to the muon anomalous magnetic moment and explain the $(g-2)_{\mu}$ anomaly~\cite{Amaral:2021rzw}. Henceforth, we refer to the $U(1)_{L{\mu}-L{\tau}}$ model, where $X_1$ only couples to the SM particles, as the ``vanilla" $U(1)_{L{\mu}-L{\tau}}$ model.
The existence of dark matter (DM) and its observed abundance is one of the greatest mysteries in physics. Recent studies have revealed that an extended $U(1)_{L{\mu}-L{\tau}}$ model, which introduces a dark matter particle with mass $M(\chi)$ and coupling to MeV-scale $X_1$ with the coupling strength $g_{D}'$, can also explain DM phenomena and the relic abundance of DM~\cite{Kamada:2018zxi,Foldenauer:2018zrz,Kahn:2018cqs,Patra:2016shz,Shuve:2014doa}. For $M(\chi)<M({X_1})/2$ and coupling ratios  $g_D'/g_{1}' \gg 1$, the dominant decay
mode of the $X_1$ is invisible, $X_1\to \chi\bar{\chi}$.
We henceforth refer to the model with $\mathcal{B}(X_1\to\chi\bar{\chi})\approx 1$ as the ``invisible" $U(1)_{L{\mu}-L{\tau}}$ model.
In addition to a vector boson scenario, an extra $U(1)$ group involving a new light scalar boson $X_0$, coupling to muons with coupling strength $g_{0}'$, has been recently addressed~\cite{Chen:2017awl,Capdevilla:2021kcf,Cvetic:2020vkk}. This model can also serve as one possible explanation for the $(g-2)_{\mu}$ discrepancy within a specific $X_0 -g_{0}'$ parameter space. In the following, this model is denoted as the ``scalar" $U(1)$ model.

Stringent constraints on the visible decay of \mbox{$X_1\to\mu^+\mu^-$} have been obtained in the BABAR~\cite{BaBar:2016sci}, CMS~\cite{CMS:2018yxg} and Belle~\cite{Belle:2021feg} experiments. 
The parameter space with $10^{-3}<g_1'<1$ in the mass range from the $\uu$ threshold to 68 GeV$/c^2$ has been excluded.
Moreover, since the ``vanilla" $U(1)_{L{\mu}-L{\tau}}$ model modifies the neutrino interactions, there are also strong constraints on the $g_{1}'$ coupling from neutrino trident $\nu N\to\nu N\uu$ scattering experiments~\cite{Altmannshofer:2014pba,Kamada:2015era,Gninenko:2020xys,AtzoriCorona:2022moj}. Therefore, $X_1$ could weakly couple to the SM particles and predominantly decay to invisible final states, especially when the mass of $X_{0,1}$ is below $2m_\mu$.
Recently, the invisible decays of $X_1$ have been investigated in the NA64-e~\cite{NA64:2022rme} and \mbox{Belle II}~\cite{Belle-II:2022yaw,Belle-II:2019qfb} experiments. There is no direct result on the scalar boson $X_0$, but it could be estimated based on the vector scheme~\cite{Capdevilla:2021kcf}.
BESIII offers significant advantages in searching for the  low-mass $X_{0,1}$ particles via the $J/\psi\to\uu X_{0,1}$ decay, where the $X_{0,1}$ is radiated from one of the muons and then decays invisibly. 
First, BESIII has collected a large $\jpsi$ data sample at the $\ee$ center-of-mass energy $\sqrt{s}=3.097$~GeV~\cite{BESIII:2021cxx}. The corresponding cross section of $\jpsi\to\uu X_{0,1}$ is approximately 22 times greater than that of the continuum $\ee\to\uu X_{0,1}$ process, which was previously employed in the search for $X_1$ at Belle II~\cite{Cvetic:2020vkk}.
Second, the lower $\ee$ collider energy at BESIII leads to a better detection resolution, enabling a finer binning scheme in search of low-mass $X_{0,1}$.
Additionally, the very narrow width of $\jpsi$ results in a lower background level from the initial state radiation process. Hence, $\jpsi\to\uu$ offers an ideal opportunity to search for muonic new physics particles~\cite{Cvetic:2020vkk}.

In this paper, we perform a search for a light muon philic scalar $X_0$ or vector $X_1$ in the processes \mbox{$J/\psi\to\uu X_{0,1}$} with $X_{0,1}\to \rm{invisible}$, in the mass range from 1 to 1000~MeV$/c^2$, based on the data sample of 9 billion $J/\psi$ events collected by the BESIII detector in 2009, 2018, and 2019~\cite{BESIII:2021cxx}. 
The data collected in 2012 is not used because information from the muon counter detectors is unavailable.
Three SM extension models, including the ``vanilla" $U(1)_{L{\mu}-L{\tau}}$ model, the ``invisible" $U(1)_{L{\mu}-L{\tau}}$ model and the ``scalar" $U(1)$ model, are considered. In the vanilla $U(1)_{L{\mu}-L{\tau}}$ model, $X_1$ decays to neutrinos with the branching fraction $\mathcal{B} (X_{0,1}\to\nu\bar{\nu}$) varying from 33$\%$ to 100$\%$ depending on the $X_1$ mass~\cite{Araki:2017wyg}. In the invisible $U(1)_{L{\mu}-L{\tau}}$ model, $X_1$ predominantly decays into light DM particles with a branching fraction $\mathcal{B}(X_1\to\chi\bar{\chi})\simeq 1$. In the scalar $U(1)$ model, $X_0$ is long-lived with displaced decay or decays predominantly to invisible particles. For all models, it is assumed that the total width $\Gamma_{X_{0,1}}$ of the $X_{0,1}$ is negligible compared to the experimental resolution and so is set to zero. Therefore, we look for events with two final state muon tracks with missing energy.
In the presence of the $X_{0,1}$ signals, narrow peaks would be visible in the recoil mass distribution of the $\uu$ system. The branching fractions of the $\jpsi\to\uu X_{0,1}$, $X_{0,1}\to \rm{invisible}$ decays are calculated as
\begin{equation}
\label{eq-bf}
\mathcal{B}(\jpsi\to\uu X_{0,1})\times\mathcal{B}(X_{0,1}\to \rm{invisible})=\frac{\it{N}_{\rm{X_{0,1}}}}{\it{N}_{\jpsi}\epsilon_{\rm{X_{0,1}}}},
\end{equation}
where $N_{X_{0,1}}$ are the signal yields, $N_{\jpsi}=(8.998\pm 0.039)\times10^9$ is the total number of $\jpsi$ events, and $\epsilon_{X_{0,1}}$ are the signal efficiencies for the $X_0$ and $X_1$ cases, respectively. The values of the coupling $g'_{0,1}$ can be obtained by converting the results for the branching fractions~\cite{Cvetic:2020vkk}.

The BESIII detector is described in detail elsewhere~\cite{Ablikim:2009aa,etof,Huang:2022wuo}. Simulated Monte Carlo (MC) samples produced with a {\sc
geant4}-based~\cite{geant4} package
are used to determine detection efficiencies
and to estimate backgrounds. 
The signal events for the $\jpsi\to\uu X_{0,1}$ with $X_{0,1}\to\rm{invisible}$ decays are generated based on the theoretical amplitude in Ref.~\cite{Cvetic:2020vkk} with {\sc
evtgen}~\cite{ref:evtgen}.
We generate signal MC samples at 58 different $X_{0,1}$ mass values with negligible width $\Gamma_{X_{0,1}}$, corresponding to mass hypotheses ranging from 1 to 1000~MeV$/c^2$ in steps of 10 to 20~MeV$/c^2$ depending on the resolution.
Possible backgrounds are investigated using an inclusive MC sample including the production of the $J/\psi$
resonance and the continuum processes.
All particle decays are modeled with {\sc
evtgen} using branching fractions either taken from the Particle Data Group~\cite{pdg}, when available,
or otherwise estimated with {\sc lundcharm}~\cite{ref:lundcharm}.  The 
background from $\ee\to\uu$ events is modeled with~ {\sc babayaga}~\cite{CarloniCalame:2003yt}.
Final state radiation is simulated with the {\sc photos} package~\cite{photos}.

The signal candidates include $\mu^+\mu^-$ tracks and missing energy in the final states. 
Each charged track is required to be within a polar angle ($\theta$) range of $|\rm{cos\theta}|<0.93$ 
and the distance of closest approach to the interaction point must be less than 10\,cm along the $z$-axis, and less than 1\,cm in the transverse plane. 
To exclude the cosmic ray events, the momentum of each track is required to be less than 1.55~GeV$/c$, and the time difference in time-of-flight system (TOF) between two tracks is required to be within $|\Delta t_{\rm{TOF}}|<2$\,ns. 
Events with additional charged tracks reconstructed in the main drift chamber (MDC) are rejected to exclude background from particles with displaced decays, such as $K_S^0$ and $\gamma$-conversion events. 
Moreover, to distinguish muons from electrons, pions, and kaons, a track is identified as a muon by the following requirements:  
(1) particle identification~(PID) likelihoods, formed by combining the measurements of the energy deposited in the MDC and the flight time in the TOF, satisfy $\mathcal{L}(\mu)>\mathcal{L}(K)$ and $\mathcal{L}(\mu)>0$, where
$\mathcal{L}(\mu)$ and $\mathcal{L}(K)$ are likelihoods calculated based on the muon and kaon hypotheses, respectively;
(2) the penetration depth of each track in the  muon counters (MUC) is required to exceed [$-$40.0 + 70 $\times\, p$/(GeV$/c$)] cm for $0.5\le p \le1.1$~GeV$/c$, and to be greater than 40~cm for $p>1.1$~GeV$/c$, where $p$ is the momentum of each charged track; 
(3) the deposited energy $E_{\rm{EMC}}(\mu)$ in the electromagnetic calorimeter (EMC) is required to be within (0.1, 0.3)~GeV.

We split the events that pass the above selections into two groups, labeled the low mass region and the high mass region.
The following selection criteria are optimized for the $X_0$ and $X_1$ searches in the low  and high mass regions, respectively.
In the low mass region, the signals are identified in the recoil mass squared $M_{\rm recoil}^2(\uu)=(p_{\jpsi}-p_{\mu^+}-p_{\mu^-})^2$ distribution since $M_{\rm recoil}^2(\uu)$ can be negative due to detector resolution, where the $p_{\jpsi}$~\cite{BESIII:2021cxx}, $p_{\mu^+}$, and $p_{\mu^-}$ are the four-momenta of $\jpsi$, $\mu^+$, and $\mu^-$ particles, respectively.
This region is defined as $M_{\rm recoil}^2(\uu)<0.3$~GeV$^2/c^4$, which is used to search for signals with mass $M(X_{0,1})<0.4$~GeV$/c^2$. The dominant backgrounds in this region are composed of two final state muons from the $J/\psi \to\mu^+\mu^-(\gamma)$ and $\ee\to\mu^+\mu^-(\gamma)$ processes, where the photons are from initial and final state radiation. The high mass region is defined as \mbox{$M_{\rm recoil}(\uu)\in[0.25,1.0]$~GeV$/c^2$} and is used to search for the signal with mass $M(X_{0,1})\in[0.4,1.0]$~GeV$/c^2$ in the recoil mass distribution. 
The region
$M_{\rm recoil}(\uu)>1.0$ GeV/$c^2$ is not studied in this work due to poor
understanding of the background.
In addition to the background from $J/\psi\to\mu^+\mu^-(\gamma)$ and $\ee\to\mu^+\mu^-(\gamma)$ processes, a significant background comes from $J/\psi\to K_L^0\pi^{\pm}K^{\mp}$ and $J/\psi\to K_L^0\pi^{+}\pi^{-}$ decays, where the $K_L^0$ particle is undetected due to its long decay length.

The common selection criteria for all events are described below. We require the events with the polar angle of the momentum of $\uu$ system $\cos(\theta_{\uu})$ within the barrel EMC region $\cos(\theta_{\uu})\in[-0.76,0.76]$, to reject the inefficient region of photon detection. The events within $\cos(\theta_{\uu})\in[-0.03,0.03]$, where photons easily escape due to the crystals being placed perpendicular to the beam direction, are further excluded. Additionally, the invariant mass of $\uu$ pairs is required to be outside the mass windows of $M(\uu)\in[0.429,0.481]$~GeV$/c^2$ and $M(\uu)\in[0.548,0.780]$~GeV$/c^2$, to further suppress the backgrounds with $K_S^0\to\pi^+\pi^-$ and $K^*\to K^{\pm}\pi^{\mp}$ decays, respectively. 
The signal processes have two final state muons; to suppress the background with additional final state photons, the total deposited energy $E_{\rm{tot}}$(EMC) in the EMC from all showers in each event is required to be less than 43~MeV (50~MeV) for the low (high) mass region. For these showers, the difference between the EMC time and the event start time is required to be within [0, 700]~ns to minimize the impact of electronic noise and showers unrelated to the event.
In each event, we require the cosine of the opening angle in the $J/\psi$ rest frame to satisfy $\cos(\alpha_{\mu^+\mu^-}) > -0.97$ (0.96) for the $X_1$ ($X_0$) case in the low mass region and $\cos(\alpha_{\mu^+\mu^-}) > -0.97$ for both cases in the high mass
region to suppress background from $J/\psi \to \mu^+\mu^-$ and $e^+
e^- \to \mu^+\mu^-$ processes. In the high mass region,
the background from $J/\psi\to K^{\pm} K^{*\mp}$ with $K^{*\mp}\to K_L^0\pi^{\mp}$ decays is suppressed by requiring the muon-hypothesis recoil mass $M_{\rm recoil}(\mu^{\pm})=|p_{\jpsi}-p_{\mu^{\pm}}|$ to be outside $(0.858, 1.172)$~GeV$/c^2$.

\begin{figure}[htpb]
	\centering
 	
	\subfigure{
  \includegraphics[trim=0 65 0 18,clip,width=0.8\linewidth]{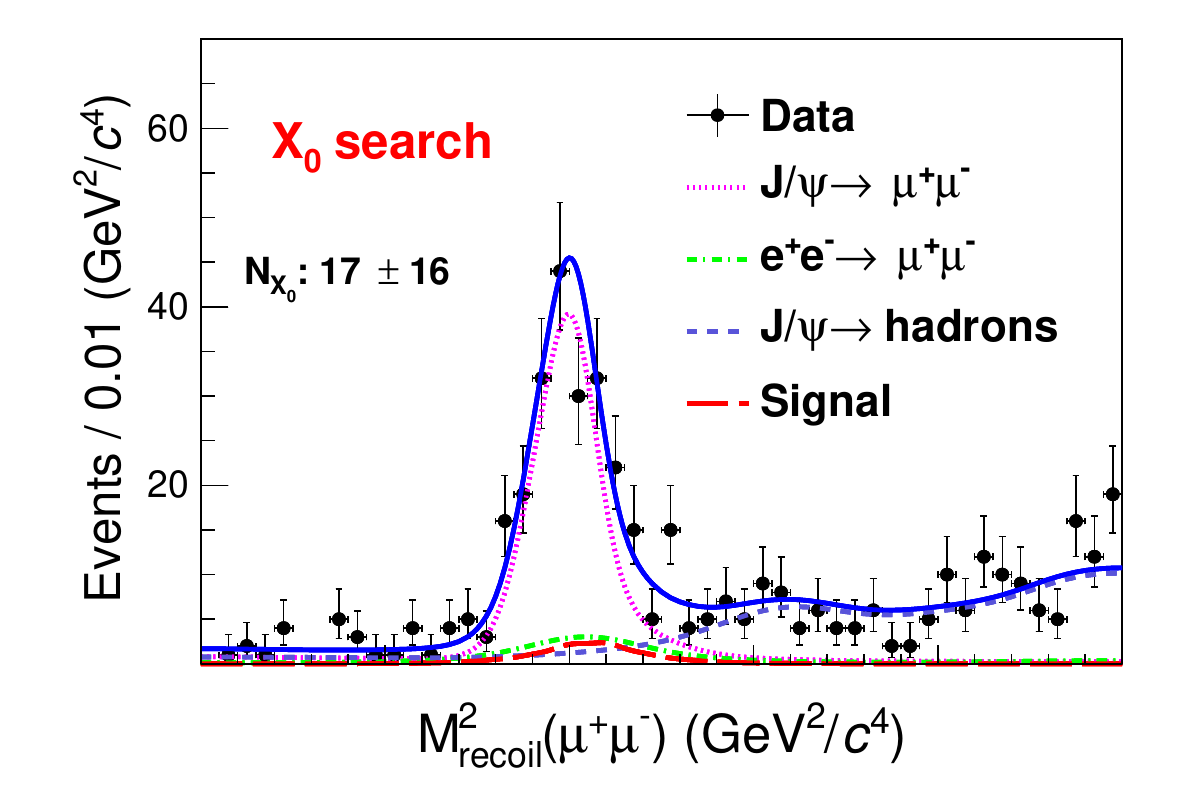}
}
\vskip -5pt
	\subfigure{
  \includegraphics[trim=0 0 0 18,clip,width=0.8\linewidth]{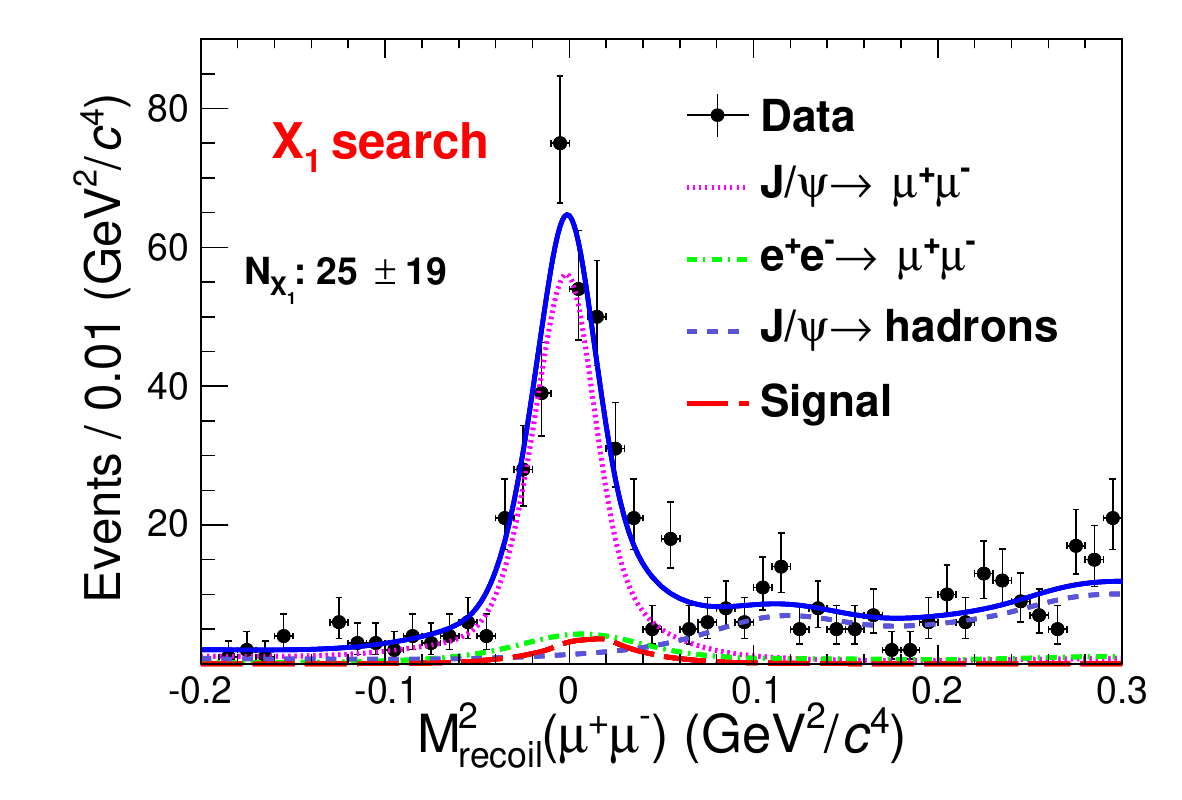}
}
\caption{Fits to the $M_{\rm recoil}^2(\uu)$ distributions for the $X_{0}$ (top) and $X_{1}$ (bottom) candidates with mass $M(X_{0,1})=120$~MeV$/c^2$. The red long-dashed curves are the $X_{0,1}$ signal shapes. The green dashed and magenta dotted curves are the peaking backgrounds from $\jpsi\to\uu, \ee\to\uu$ processes. Dots with error bars are data, and the blue dashed curves are the backgrounds from $\jpsi\to\rm{hadrons}$ processes.}
 	\label{fig:fit-R1}	
\end{figure}

\begin{figure}[htpb]
	\centering
	\subfigure{
  \includegraphics[trim=0 65 0 18,clip,width=0.8\linewidth]{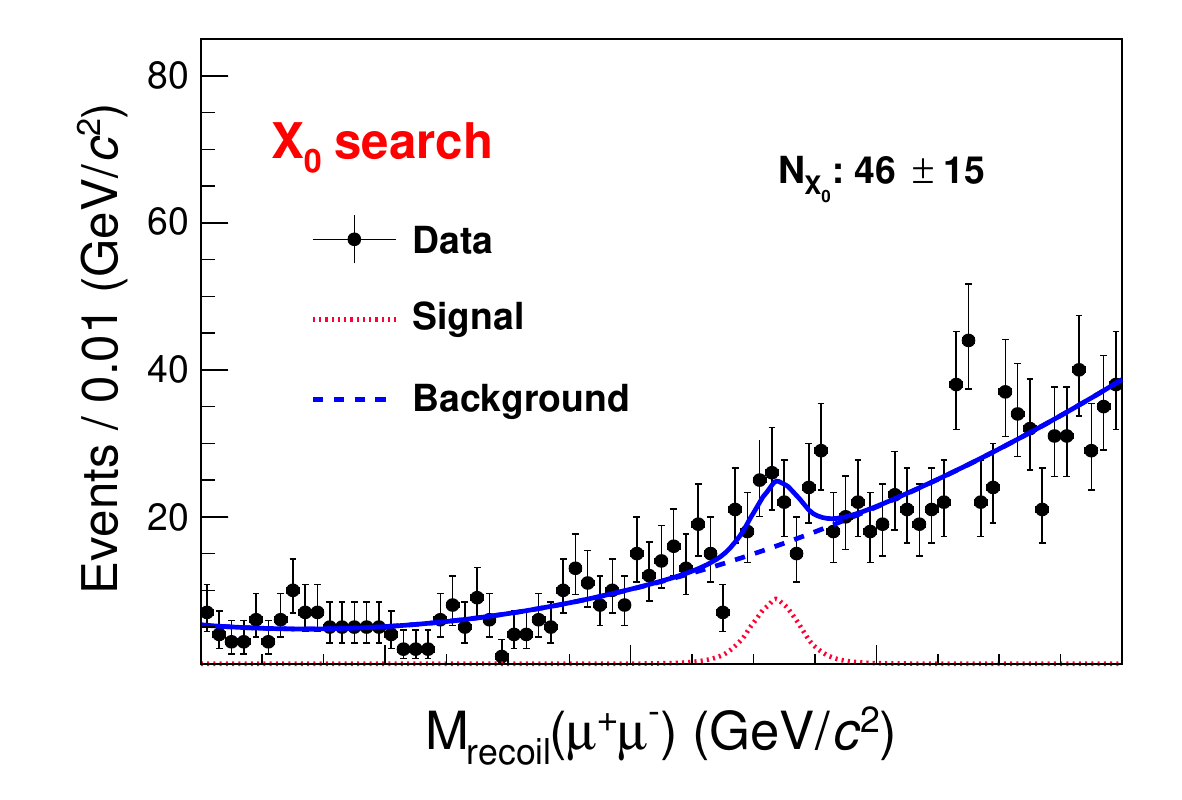}
}
\vskip -5pt
	\subfigure{
  \includegraphics[trim=0 0 0 18,clip,width=0.8\linewidth]{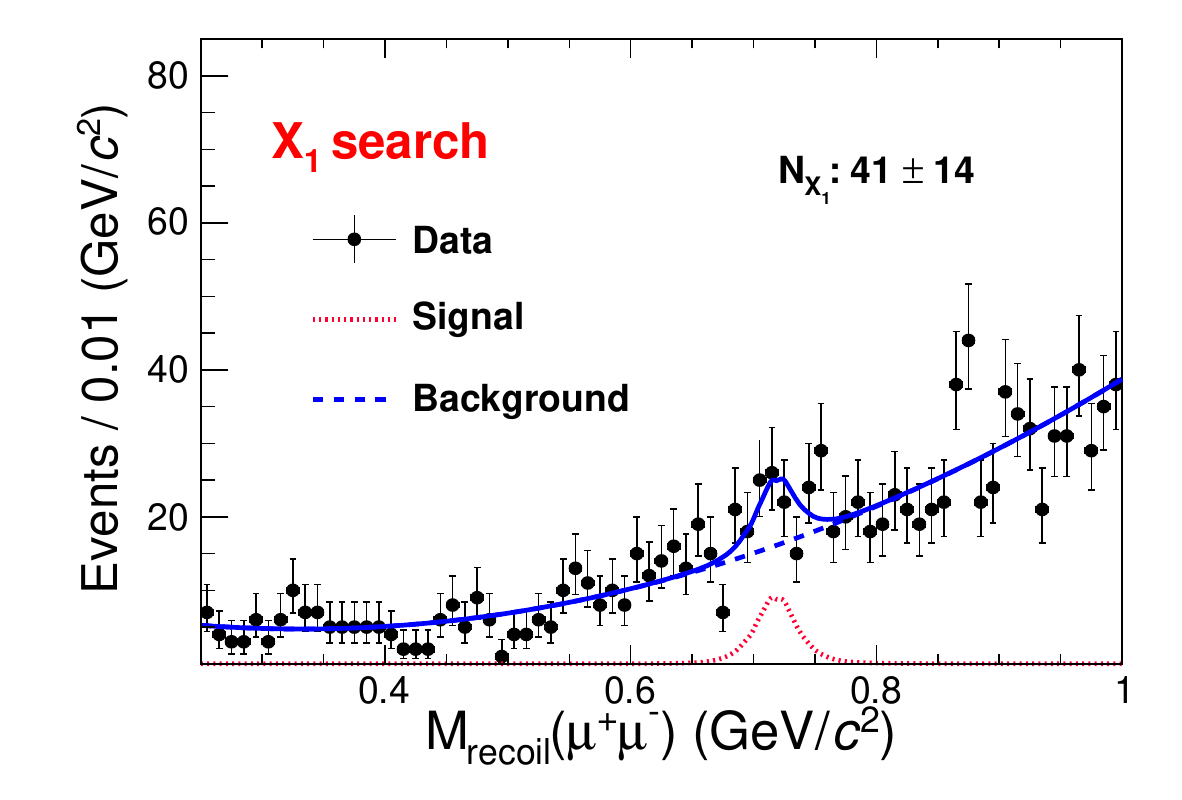}
}
 
 \caption{Fits to the $M_{\rm recoil}(\uu)$ distributions for the $X_{0}$ (top) and $X_{1}$ (bottom) candidates with mass $M(X_{0,1})=720$~MeV$/c^2$. The red dotted curves are the $X_{0,1}$ signal shapes. Dots with error bars are data, and the blue dashed curves describe the combinatorial background contribution. The data points on the two plots are identical due to identical selection criteria for $X_0$ and $X_1$. }
	\label{fig:fit-R2}	
\end{figure}

To investigate possible signals from $X_{0,1}$, we perform a series of unbinned maximum likelihood fits to $M_{\rm recoil}^2(\uu)$ for the low mass region and to $M_{\rm recoil}(\uu)$ for the high mass region. 
The fits are performed in the mass region of $1<M(X_{0,1})<1000$~MeV$/c^2$, with steps of $10-20$~MeV$/c^2$~(about half of the signal resolution).
The resulting signal efficiency obtained from signal MC samples varies between $1\% - 20\%$ as a function of $M(X_{0,1})$.
In the low mass region, 
the residual backgrounds are from $\jpsi\to\uu\gamma$, $\ee\to\uu\gamma$, and $\jpsi\to\rm{hadrons}$ processes. The yields and probability density functions for the peaking backgrounds from $\jpsi\to\uu$ and $\ee\to\uu$ are fixed from the corresponding MC simulation in these fits. The shape of non-peaking background from $\jpsi$ hadronic decays is constructed from the corresponding MC simulation with a yield that is left as a free parameter in the fit. The signal shapes are described as a templated shape constructed from the $X_{0,1}$ signal MC simulations. As an example, Fig.~\ref{fig:fit-R1} shows the fit results for the $X_{0,1}$ candidates with mass $M(X_{0,1})=120$~MeV$/c^2$. 
For the high mass region, the signals with mass $M(X_{0,1})>0.4$~GeV$/c^2$ are searched for in the $M_{\rm recoil}(\uu)$ distribution, and the residual backgrounds are dominated by the $J/\psi$ decays with $K_L^{0}$ in the final state, which are described by a second-order Chebychev function. The signal shapes are constructed from the corresponding $X_{0,1}$ signal MC simulations. The fits to the $M_{\rm recoil}(\uu)$ distributions with mass $M(X_{0,1})=720$~MeV$/c^2$ are shown in Fig.~\ref{fig:fit-R2}. Taking into account the uncertainty from the background model, the maximum local significance among all the fits is determined to be 2.5$\sigma$ at $M(X_{0,1})=720$~MeV$/c^2$, as shown in Fig.~\ref{fig:fit-R2}. The significances are calculated by comparing the likelihoods with and without the signal components in the fit, and considering the change of the number of degrees of freedom. We find no evidence for signals from $X_{0,1}$ invisible decays.

\begin{figure*}[htpb]
         \textbf{``vanilla" $L_{\mu}-L_{\tau}$ model\,\,\,\,\,\,\, \,\,\,\,\,\,\,\,\, \,\,\,\,\,\ \,\,\, ``invisible" $L_{\mu}-L_{\tau}$ model \,\,\,\,\,\,\, \,\,\,\,\,\,\,\,\, \,\,\,\,\,\ \,\,\,   ``scalar" $U(1)$ model }
  	\subfigure{
		\includegraphics[width=2.3in,height=2.0in]{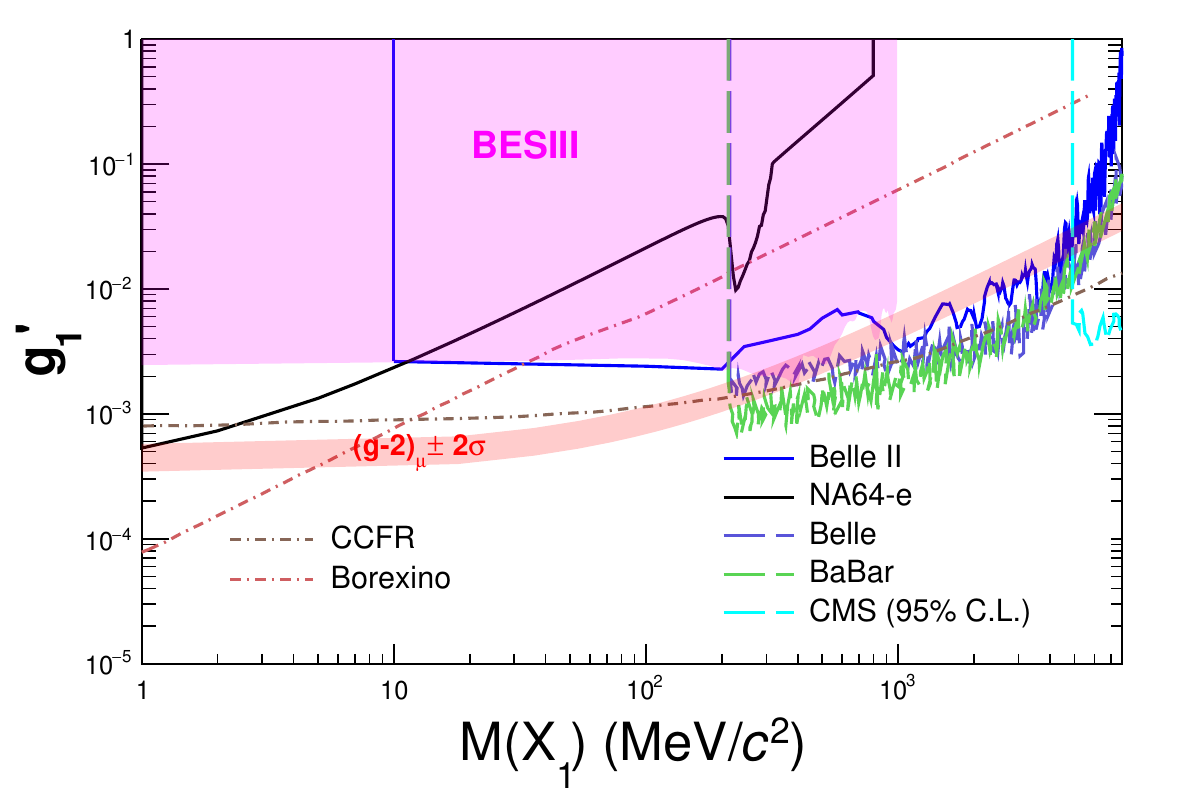}
  	\includegraphics[width=2.3in,height=2.0in]{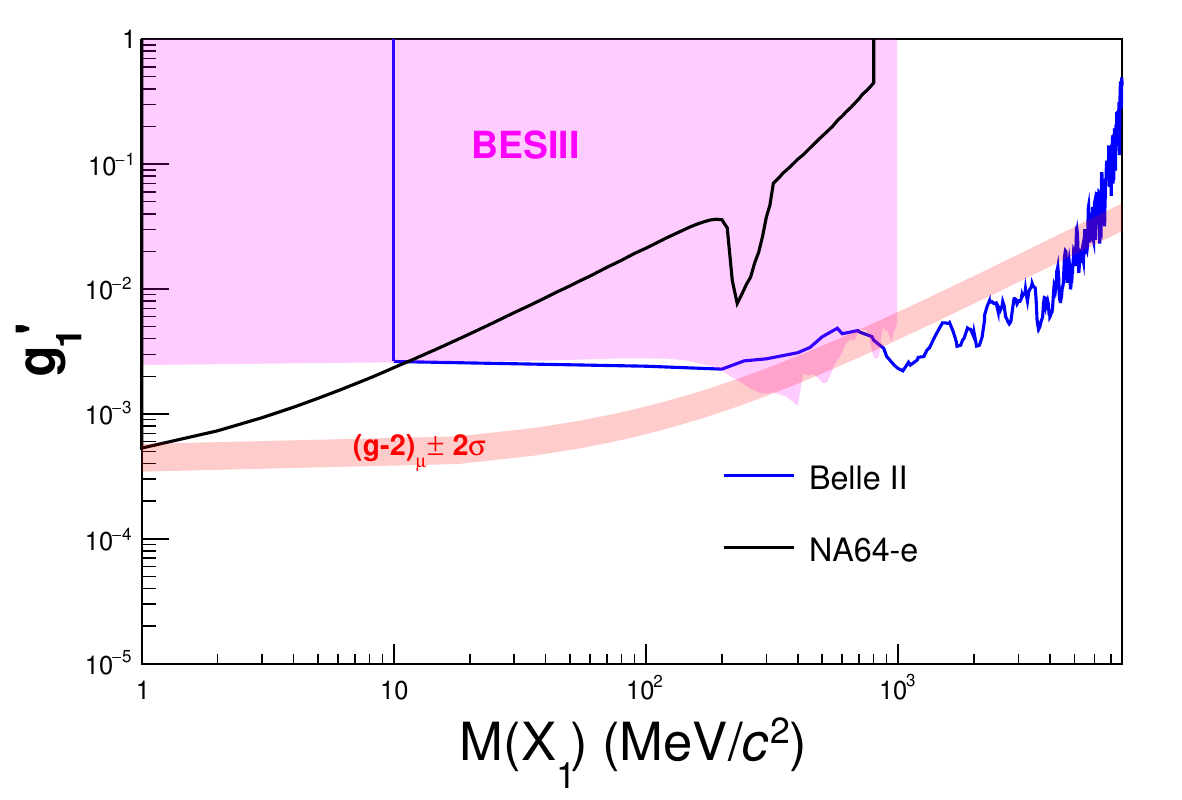}
   \includegraphics[width=2.3in,height=2.0in]{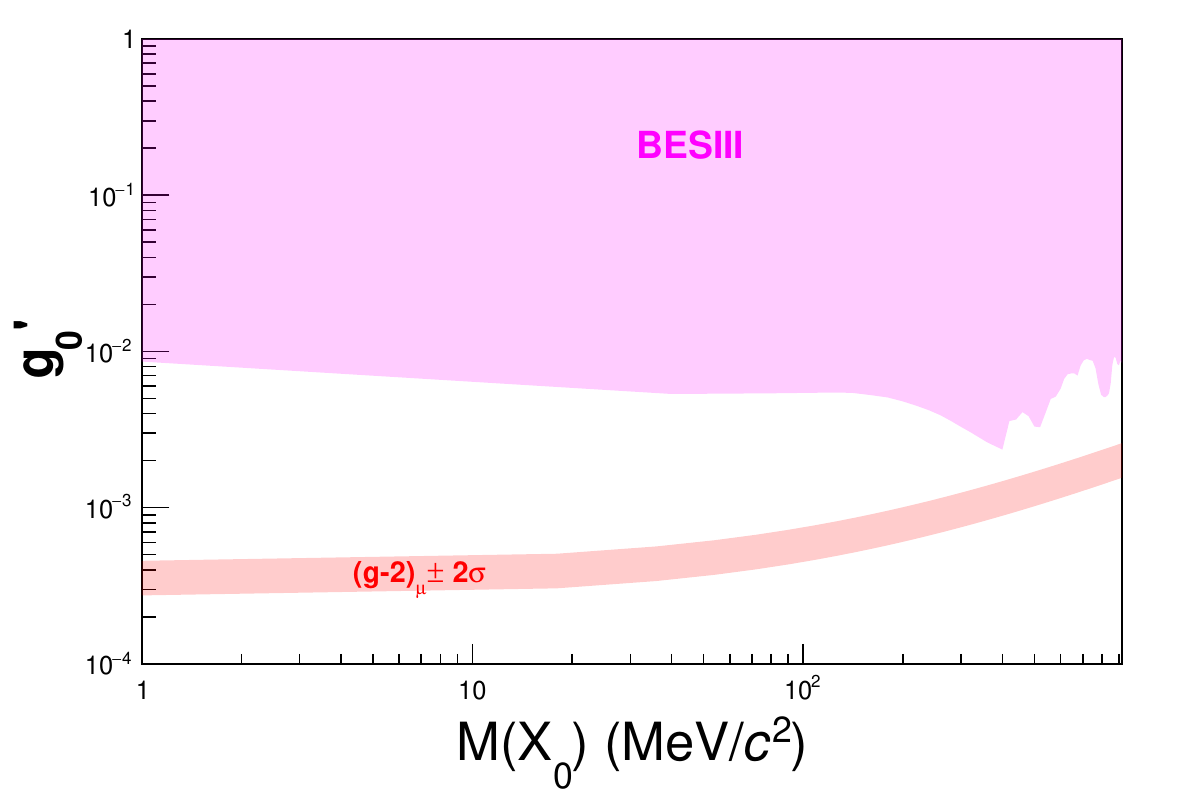}
	}    				
 	\caption{The 90$\%$ C.L. upper limits on the coupling $g_{0,1}'$ for the vanilla $L_{\mu}-L_{\tau}$ model, the invisible $L_{\mu}-L_{\tau}$ model, and the ``scalar" $U(1)$ model. For the ``vanilla" $L_{\mu}-L_{\tau}$ model, the previous excluded regions by BABAR~\cite{BaBar:2016sci}, CMS~\cite{CMS:2018yxg}, and Belle~\cite{Belle:2021feg} via the $X_{1}\to\uu$  decay and the constraints from neutrino experiments~\cite{Altmannshofer:2014pba,Kamada:2015era,Gninenko:2020xys,AtzoriCorona:2022moj} are shown for comparison. 
    For the invisible $L_{\mu}-L_{\tau}$ model, the previous searches for $X_1$ invisible decays by NA64-e~\cite{NA64:2022rme}  and Belle II~\cite{Belle-II:2022yaw,Belle-II:2019qfb} are also presented. 
    The red bands represent the parameter regions favored by the $(g-2)_{\mu}$ anomaly within $2\sigma$ \cite{Cvetic:2020vkk}.}	
	\label{fig:ul-g}	
\end{figure*} 

The systematic uncertainty sources for the coupling $g'_{0,1}$ measurement include the total number of $\jpsi$ events, the signal efficiency,  the signal extraction, and the $\jpsi$ total width. The uncertainty from the total number of $\jpsi$ events is 0.4$\%$~\cite{BESIII:2021cxx}. 
 The uncertainty from the tracking efficiency is taken as 1.0$\%$ per track~\cite{BESIII:2021ges}. The uncertainty associated with the $E_{\rm{EMC}}(\mu)$ requirement is 0.2$\%$~\cite{BESIII:2021ges}. 
 The systematic uncertainties from the requirements on $\Delta t_{\rm{TOF}}$, $E_{\rm{tot}}(\rm{EMC})$ and additional tracks in the MDC are estimated with a $\jpsi\to\uu$ control sample. The resulting differences in the efficiencies between data and MC simulation are 1.0$\%$ for $\Delta t_{\rm{TOF}}$, 1.5$\%$ for $E_{\rm{tot}}(\rm{EMC})$, and negligible for the requirement on the additional tracks.
The uncertainties due to the requirements on PID and MUC penetration depth are investigated with a $\jpsi\to\gamma\uu$ control sample with tagged photon. The signal MC events are weighted by the difference in efficiency between data and MC simulation. It causes $1.0\%$ and $(1.0 - 3.0)\%$ changes in signal efficiencies for PID and penetration depth requirement depending on the mass of $X_{0,1}$. In the fits for the low mass region, 
the uncertainty from the simulated yields of the peaking background is estimated to be 18.8$\%$ and 16.8$\%$ for $X_0$ and $X_1$, respectively, using the selected $\jpsi\to\gamma\uu$ control sample.
To account for these uncertainties, alternative fits are performed by varying the background yields upward or downward by their corresponding uncertainties.
For the high mass region, the uncertainty associated with the background model is also considered by an alternative fit with the background described by a third-order polynomial shape. Among these fits, the one with the largest upper limit on the signal yield is treated as the final result. The systematic uncertainty from the resolution difference between data and MC simulation is determined to be 1.0$\%$ using a control sample of $\jpsi\to K_{S}^{0}K_{L}^0$ events. Considering that the cross section of the $\ee\to\uu$ process is 4.4$\%$ of the $\jpsi\to\ee$ one~\cite{Asner:2008nq}, and taking into account the efficiency difference between $\ee \to\uu X_{0,1}$ and $\jpsi\to\uu X_{0,1}$, the uncertainties for the contributions from the $\ee \to\uu X_{0}$ and $\ee \to\uu X_{1}$ processes are determined to be  3.5$\%$ and $3.9\%$, respectively.
The uncertainty associated with the total width of $\jpsi$ is 1.8$\%$~\cite{pdg}. 
Assuming all these sources as independent, the total systematic uncertainty is obtained by adding the individual contributions in quadrature, resulting in $(5.0-5.4)\%$ and $(5.5-5.8)\%$ for $X_{0}$ and $X_{1}$, respectively.


Since no obvious signal is observed, the upper limits on the product branching fractions $\mathcal{B}(\jpsi\to\uu X_{0,1})\times\mathcal{B}(X_{0,1}\to \rm{invisible})$ in Eq.~(\ref{eq-bf}) are determined at the 90$\%$ credibility level (C.L.) depending on the mass $M(X_{0,1})$ with Bayesian method.
We perform a likelihood scan by conducting a series of fits under different hypotheses for the number of $X_{0,1}$.
The effect of the systematic uncertainty is considered by convolving the likelihood curve with a Gaussian function with its standard deviation set to the total systematic uncertainty~\cite{Stenson:2006gwf,Liu:2015uha}. The resulting 90$\%$ C.L. upper limits on $\mathcal{B}(\jpsi\to\uu X_{0,1})\times\mathcal{B}(X_{0,1}\to \rm{invisible})$ are determined to be $6.2\times10^{-9} - 5.5\times10^{-7}$ and $4.5\times10^{-9} - 9.6\times10^{-7}$ for the cases of $X_{0}$ and $X_{1}$ as functions of $M(X_{0,1})$, respectively. The results on the branching fractions are used to estimate limits on the coupling $g_{0,1}'$~\cite{Cvetic:2020vkk}. The excluded region in the $g_{0,1}'$ versus $M(X_1) $  parameter space at the 90$\%$ C.L. is shown in Fig.~\ref{fig:ul-g}. For the vanilla $L_{\mu}-L_{\tau}$ model, BESIII excludes $g_1'$ values in the range $1.6\times 10^{-3} - 7.9\times 10^{-3}$ as a function of $M(X_1)$ after taking $\mathcal{B}(X_{1}\to\nu\bar{\nu})$ into account. 
For the invisible scenario, $g_1'$ values in the range $1.1\times 10^{-3} - 5.5\times 10^{-3}$ with $1 \le  M(X_1)\le 1000$~MeV$/c^2$ are excluded.
 We obtain a better sensitivity in the range 200 - 860~MeV$/c^2$ compared to the Belle II results~\cite{Belle-II:2022yaw,Belle-II:2019qfb}, and a comparable upper limit in the lower mass region with a finer binning scheme. The best constraint for the mass region $M(X_1)<10$~MeV$/c^2$ is provided in the NA64-e experiment~\cite{NA64:2022rme}. For the ``scalar" $X_0$ case, there are no earlier experimental measurements; the 90$\%$ C.L. upper limits on the coupling $g_0'$ is determined to be $2.3\times 10^{-3} - 1.0\times 10^{-2}$ with $M(X_0)$ in the range 1 - 1000~MeV$/c^2$.

In summary, we have searched for a muon philic scalar $X_{0}$ or vector $X_1$ boson, introduced by many SM extension models, using a data sample of 9 billion $\jpsi$ events at BESIII. No evidence of $X_{0,1}$ has been observed in the mass range $1  <M(X_{0,1})<1000$~MeV$/c^2$, and the 90$\%$ C.L. upper limits on the coupling $g_{0,1}'$ are set to be in the range of $1.1\times 10^{-3} - 1.0\times 10^{-2}$ for three SM extension models.
To date, we provide the best constraint for the $X_1$ mass in the range 200 - 860 MeV$/c^2$ in the invisible $L_{\mu}-L_{\tau}$ model, and two mass regions within $320< M(X_1)< 410$~MeV$/c^2$ and $460< M(X_1)< 520$~MeV$/c^2$ are excluded to explain the $(g-2)_{\mu}$ anomaly at the 90$\%$ C.L..
For the scalar $X_{0}$, we have performed the first direct experiment search, and set the upper limit at the 90$\%$ C.L. on the coupling $g_0'$ to $2.3\times 10^{-3} - 1.0\times 10^{-2}$ for $1<M(X_0)<1000$~MeV$/c^2$.

The BESIII Collaboration thanks the staff of BEPCII and the IHEP computing center for their strong support. This work is supported in part by National Key R\&D Program of China under Contracts No. 2020YFA0406400, No. 2020YFA0406300; National Natural Science Foundation of China (NSFC) under Contracts No. 11635010, No. 11735014, No. 11835012, No. 11935015, No. 11935016, No. 11935018, No. 11961141012, No. 12025502, No. 12035009, No. 12035013, No. 12061131003, No. 12192260, No. 12192261, No. 12192262, No. 12192263, 12192264, No. 12192265, No. 12221005, 12225509, No. 12235017; the Chinese Academy of Sciences (CAS) Large-Scale Scientific Facility Program; the CAS Center for Excellence in Particle Physics (CCEPP); Joint Large-Scale Scientific Facility Funds of the NSFC and CAS under Contract No. U1832207; CAS Key Research Program of Frontier Sciences under Contracts No. QYZDJ-SSW-SLH003, No. QYZDJ-SSW-SLH040; 100 Talents Program of CAS; The Institute of Nuclear and Particle Physics (INPAC) and Shanghai Key Laboratory for Particle Physics and Cosmology; European Union's Horizon 2020 research and innovation programme under Marie Sklodowska-Curie grant agreement under Contract No. 894790; German Research Foundation DFG under Contracts Nos. 455635585, Collaborative Research Center CRC 1044, FOR5327, GRK 2149; Istituto Nazionale di Fisica Nucleare, Italy; Ministry of Development of Turkey under Contract No. DPT2006K-120470; National Research Foundation of Korea under Contract No. NRF-2022R1A2C1092335; National Science and Technology fund of Mongolia; National Science Research and Innovation Fund (NSRF) via the Program Management Unit for Human Resources \& Institutional Development, Research and Innovation of Thailand under Contract No. B16F640076; Polish National Science Centre under Contract No. 2019/35/O/ST2/02907; The Swedish Research Council; U. S. Department of Energy under Contract No. DE-FG02-05ER41374.



\begin{thebibliography}{99}


\bibitem{Crivellin:2021sff}
A.~Crivellin and M.~Hoferichter,
Science \textbf{374}, 1051 (2021).



\bibitem{Fox:2022tzz}
P.~J.~Fox \textit{et al.},
arXiv:2210.03075.

\bibitem{Chen:2021fcb}
S.~Chen and S.~L.~Olsen,
Natl. Sci. Rev. \textbf{8}, nwab189 (2021).

\bibitem{Muong-2:2006rrc}
G.~W.~Bennett \textit{et al.} (Muon g-2 Collaboration),
Phys. Rev. D \textbf{73}, 072003 (2006).

\bibitem{Muong-2:2021ojo}
B.~Abi \textit{et al.} (Muon g-2 Collaboration),
Phys. Rev. Lett. \textbf{126}, 141801 (2021).


\bibitem{Muong-2:2023cdq}
D.~P.~Aguillard \textit{et al.} (Muon g-2),
Phys. Rev. Lett. \textbf{131}, 161802 (2023).



\bibitem{Pospelov:2008zw}
M.~Pospelov,
Phys. Rev. D \textbf{80}, 095002 (2009).

\bibitem{Bauer:2018onh}
M.~Bauer, P.~Foldenauer, and J.~Jaeckel,
J. High Energy Phys. \textbf{07} (2018) 094.

\bibitem{Foot:1990mn}
R.~Foot,
Mod. Phys. Lett. A \textbf{06}, 527 (1991).

\bibitem{He:1990pn}
X.~G.~He, G.~C.~Joshi, H.~Lew, and R.~R.~Volkas,
Phys. Rev. D \textbf{43}, R22 (1991).

\bibitem{Foot:1994vd}
R.~Foot, X.~G.~He, H.~Lew, and R.~R.~Volkas,
Phys. Rev. D \textbf{50}, 4571 (1994).

\bibitem{Amaral:2021rzw}
D.~W.~P.~Amaral, D.~G.~Cerdeno, A.~Cheek, and P.~Foldenauer,
Eur. Phys. J. C \textbf{81}, 861 (2021).


\bibitem{Kamada:2018zxi}
A.~Kamada, K.~Kaneta, K.~Yanagi, and H.~B.~Yu,
J. High Energy Phys. \textbf{06}, (2018) 117.

\bibitem{Foldenauer:2018zrz}
P.~Foldenauer,
Phys. Rev. D \textbf{99}, 035007 (2019).

\bibitem{Kahn:2018cqs}
Y.~Kahn, G.~Krnjaic, N.~Tran, and A.~Whitbeck,
J. High Energy Phys. \textbf{09} (2018) 153.


\bibitem{Patra:2016shz}
S.~Patra, S.~Rao, N.~Sahoo and N.~Sahu,
Nucl. Phys. B \textbf{917}, 317-336 (2017).

\bibitem{Shuve:2014doa}
B.~Shuve and I.~Yavin,
Phys. Rev. D \textbf{89}, 113004 (2014).

\bibitem{Chen:2017awl}
C.~Y.~Chen, M.~Pospelov, and Y.~M.~Zhong,
Phys. Rev. D \textbf{95}, 115005 (2017).

\bibitem{Capdevilla:2021kcf}
R.~Capdevilla, D.~Curtin, Y.~Kahn, and G.~Krnjaic,
J. High Energy Phys. \textbf{04} (2022) 129.

\bibitem{Cvetic:2020vkk}
G.~Cveti\v{c}, C.~S.~Kim, D.~Lee and D.~Sahoo,
J. High Energy Phys. \textbf{10} (2020) 207.


\bibitem{BaBar:2016sci}
J.~P.~Lees \textit{et al.} (BABAR Collaboration),
Phys. Rev. D \textbf{94}, 011102 (2016).

\bibitem{CMS:2018yxg}
A.~M.~Sirunyan \textit{et al.} (CMS Collaboration),
Phys. Lett. B \textbf{792}, 345 (2019).

\bibitem{Belle:2021feg}
T.~Czank \textit{et al.} (Belle Collaboration),
Phys. Rev. D \textbf{106}, 012003 (2022).


\bibitem{Altmannshofer:2014pba}
W.~Altmannshofer, S.~Gori, M.~Pospelov, and I.~Yavin,
Phys. Rev. Lett. \textbf{113}, 091801 (2014).

\bibitem{Kamada:2015era}
A.~Kamada and H.~B.~Yu,
Phys. Rev. D \textbf{92}, 113004 (2015).


\bibitem{Gninenko:2020xys}
S.~Gninenko and D.~Gorbunov,
Phys. Lett. B \textbf{823}, 136739 (2021).

\bibitem{AtzoriCorona:2022moj}
M.~Atzori Corona, M.~Cadeddu, N.~Cargioli, F.~Dordei, C.~Giunti, Y.~F.~Li, E.~Picciau, C.~A.~Ternes and Y.~Y.~Zhang,
J. High Energy Phys. \textbf{05} (2022) 109.




\bibitem{NA64:2022rme}
Y.~M.~Andreev \textit{et al.} (NA64 Collaboration),
Phys. Rev. D \textbf{106},  032015 (2022).

\bibitem{Belle-II:2019qfb}
I.~Adachi \textit{et al.} (Belle II Collaboration),
Phys. Rev. Lett. \textbf{124}, 141801 (2020).


\bibitem{Belle-II:2022yaw}
I.~Adachi \textit{et al.} (Belle II Collaboration),
Phys. Rev. Lett. \textbf{130}, 231801 (2023).

\bibitem{BESIII:2021cxx}
M.~Ablikim \textit{et al.} (BESIII Collaboration),
Chin. Phys. C \textbf{46}, 074001 (2022).

\bibitem{Araki:2017wyg}
T.~Araki, S.~Hoshino, T.~Ota, J.~Sato, and T.~Shimomura,
Phys. Rev. D \textbf{95}, 055006 (2017).


\bibitem{Ablikim:2009aa}
  M.~Ablikim {\it et al.} (BESIII Collaboration),
  Nucl.\ Instrum.\ Methods Phys.\ Res., Sect. A {\bf 614}, 345 (2010).

  

  
\bibitem{etof}
 X.~Li {\it et al.}, Radiat. Detect. Technol. Methods {\bf 1}, 13 (2017);
 Y.~X.~Guo {\it et al.}, Radiat. Detect. Technol. Methods {\bf 1}, 15 (2017);
 P.~Cao {\it et al.}, Nucl.\ Instrum.\ Methods Phys.\ Res., Sect. A {\bf 953}, 163053 (2020).

\bibitem{Huang:2022wuo}
K.~X.~Huang {\it et al.},
Nucl. Sci. Tech. \textbf{33}, 142 (2022).

  
\bibitem{geant4}
  S.~Agostinelli {\it et al.} [GEANT4 Collaboration],
  Nucl.\ Instrum.\ Methods Phys.\ Res., Sect. A {\bf 506}, 250 (2003).


\bibitem{ref:evtgen}
  D.~J.~Lange,
  Nucl.\ Instrum.\ Methods Phys.\ Res., Sect. A {\bf 462}, 152 (2001);
  R.~G.~Ping,
  Chin. Phys. C {\bf 32}, 599 (2008).

\bibitem{pdg}
R.~L.~Workman {\it et al.} (Particle Data Group), Prog.\ Theor.\ Exp.\ Phys. 2022, 083C01 (2022).


\bibitem{ref:lundcharm}
  J.~C.~Chen, G.~S.~Huang, X.~R.~Qi, D.~H.~Zhang, and Y.~S.~Zhu,
  Phys.\ Rev.\ D {\bf 62}, 034003 (2000);
  R.~L.~Yang, R.~G.~Ping and H.~Chen,
  Chin.\ Phys.\ Lett.\  {\bf 31}, 061301 (2014).


\bibitem{CarloniCalame:2003yt}
C.~M.~Carloni Calame, G.~Montagna, O.~Nicrosini, and F.~Piccinini,
Nucl. Phys. B, Proc. Suppl. \textbf{131}, 48 (2004).

\bibitem{photos}
  E.~Richter-Was,
  Phys.\ Lett.\ B {\bf 303}, 163 (1993).



\bibitem{BESIII:2021ges}
M.~Ablikim \textit{et al.} (BESIII Collaboration),
Phys. Rev. D \textbf{105}, 012008 (2022).

\bibitem{Asner:2008nq}
D.~M.~Asner \textit{et al.},
Int. J. Mod. Phys. A \textbf{24}, S1 (2009).


\bibitem{Stenson:2006gwf}
K.~Stenson,
arXiv:physics/0605236.


\bibitem{Liu:2015uha}
X.~X.~Liu, X.~R.~L\"u, and Y.~S.~Zhu,
Chin. Phys. C \textbf{39}, 103001 (2015).

\end{thebibliography}
\end{document}